\documentclass[final,a4paper,onecolumn]{article}
\usepackage{amsmath}

\newtheorem{theorem}{Theorem}
\newtheorem{acknowledgement}[theorem]{Acknowledgement}

\input{tcilatex}

\begin{document}

\title{Entanglement Produced by Using Some Biparticle Bose Systems}
\author{Xian-Ting Liang \\
Department of Physics and Institute of Mathematics, \\
Huaihua College, Huaihua, Hunan 418008, China\\
and\\
Department of Material Science and Engineering, University\\
of Science and Technology of China, Hefei, Anhui 230026, China}
\maketitle

\begin{abstract}
In this paper, the generating of entanglement by using some biparticle Bose
systems acting on vacuum state are investigated. These systems include
two-mode squeezed system, thermal system of a free single particle (where
the fictitious tilde system is regarded another particle), and the system of
two coupling harmonic oscillators. The technique of integration within an
ordered product (IWOP) of operators is used.


\end{abstract}

Entanglement has been recognized as a central resource in various aspects of
quantum information processing. The quantum teleportation \cite%
{teleportation}, quantum cryptography \cite{cryptography}, quantum dense
coding \cite{densecoding} etc. are all based on the entanglement. So
producing entanglement of quantum systems is one of the basic tasks for
quantum communication and quantum computation. It has been shown that
nonlocal operations can generate entanglement. Recently, many authors have
studied the generating of entanglement from simple quantum gates, for
example, CNOT gate, double CNOT gate and SWAP gate \cite{gates}. The
produced entanglement from these quantum gates is the entanglement of qubit
systems. However, recently, quantum information protocols have extended from
qubit systems to continuous variable systems, such as continuous variable
teleportation, \cite{vteleportation} continuous variable quantum computation %
\cite{vcomputation} and error correction \cite{errorcorrection}, and
continuous variable cryptography \cite{vcryptography}. Thus, it is
interested to produce entanglement from continuous variable quantum systems.
In this paper, we will try to study the generating of entanglement from some
non-qubit Bose systems by using the technique of integration within an
ordered product (IWOP) of operators. Follows the original works of Fan \cite%
{IWOPfan01} we firstly list the rules of IWOP of operators as follows.

(i) The order of Bose operators within a normally ordered product (denoted
by symbol : :) can be permuted.

(ii) If and only if the normally ordered form of operators can be put into
and out of the symbol : : unchangeably.

(iii) C numbers can be taken into and out of the symbol : : as one wishes.

(iv) A normally ordered product can be integrated or differentiated with
respect to C-number provided the integration is convergent.

(v) The normally ordered form of projector of vacuum state is 
\begin{equation}
\left| 0\right\rangle \left\langle 0\right| =:e^{-a^{\dagger }a}:,
\label{Eq1}
\end{equation}%
where $a^{\dagger }$ and $a$ are creation and annihilation operators of Bose
system in Fock space.

At first, we study the generating of entanglement by two-mode squeezed
operator acting on vacuum state. The scheme for generating two-mode squeezed
operator has been studied in last years and recently by using a pair distant
cavities \cite{squeezed}. Because the two-mode squeezed operator is $%
e^{\lambda \left( a_{1}^{\dagger }a_{2}^{\dagger }-a_{1}a_{2}\right) }$ the
created entangled state from vacuum state is just the two-mode squeezed
vacuum state%
\begin{equation}
\left| \psi \right\rangle =e^{\lambda \left( a_{1}^{\dagger }a_{2}^{\dagger
}-a_{1}a_{2}\right) }\left| 00\right\rangle ,  \label{Eq2}
\end{equation}%
where $\lambda $ is the squeezing parameter, $a_{1}^{\dagger }$ $\left(
a_{1}\right) $ and $a_{2}^{\dagger }$ $\left( a_{2}\right) $ are the
creation and (annihilation) operators of the two particles. The degree of
entanglement of $\left| \psi \right\rangle $ has been given by many authors %
\cite{squeezed}, \cite{squeezed1} \cite{squeezed2}. Where we investigate it
with IWOP of operators. It is shown that the normally ordered form of
two-mode squeezed operator is \cite{Fantwomode}%
\begin{eqnarray}
\left| \psi \right\rangle  &=&\sec h\lambda   \notag \\
&:&\left\{ \exp \left( a_{1}^{\dagger }a_{2}^{\dagger }-a_{1}a_{2}\right)
tanh\lambda \right\} \exp \left\{ \left( a_{1}^{\dagger
}a_{1}+a_{2}^{\dagger }a_{2}\right) \left( \sec h\lambda -1\right) \right\}
:\left| 00\right\rangle ,  \notag \\
&&  \label{Eq3}
\end{eqnarray}%
so we have%
\begin{equation}
\rho _{12}^{tms}=\sec h^{2}\lambda \exp \left( a_{1}^{\dagger
}a_{2}^{\dagger }tanh\lambda \right) \left| 00\right\rangle \left\langle
00\right| \exp \left( a_{1}a_{2}tanh\lambda \right) .  \label{Eq4}
\end{equation}%
where and in the following we use the superscript ``tms'' denoting
``two-mode squeezed''. The reduce density matrix of $\rho _{12}^{tms}$ is%
\begin{equation}
\rho _{1}^{tms}=tr_{2}\left\{ \sec h^{2}\lambda \exp \left( a_{1}^{\dagger
}a_{2}^{\dagger }tanh\lambda \right) \left| 00\right\rangle \left\langle
00\right| \exp \left( a_{1}a_{2}tanh\lambda \right) \right\} .  \label{Eq5}
\end{equation}%
By using the completeness of coherent state $\left| z_{2}\right\rangle $ 
\begin{equation}
\int \frac{d^{2}z_{2}}{\pi }\left| z_{2}\right\rangle \left\langle
z_{2}\right| =1,  \label{Eq6}
\end{equation}%
we have%
\begin{eqnarray}
&&\rho _{1}^{tms}  \notag \\
&=&tr_{2}\{\int \frac{d^{2}z_{2}}{\pi }\left| z_{2}\right\rangle
\left\langle z_{2}\right| \sec h^{2}\lambda \exp \left( a_{1}^{\dagger
}a_{2}^{\dagger }tanh\lambda \right) \left| 00\right\rangle \left\langle
00\right| \exp \left( a_{1}a_{2}tanh\lambda \right) \}  \notag \\
&=&\int \frac{d^{2}z_{2}}{\pi }\left\langle z_{2}\right| \sec h^{2}\lambda
\exp \left( a_{1}^{\dagger }a_{2}^{\dagger }tanh\lambda \right) :\left|
00\right\rangle \left\langle 00\right| :\exp \left( a_{1}a_{2}tanh\lambda
\right) \left| z_{2}\right\rangle   \notag \\
&=&\int \frac{d^{2}z_{2}}{\pi }\sec h^{2}\lambda \exp \left( a_{1}^{\dagger
}z_{2}^{\ast }tanh\lambda \right) :\left| 0\right\rangle \left\langle
0\right| :\exp \left( a_{1}z_{2}tanh\lambda \right) \left| \left\langle
0\right| \left. z_{2}\right\rangle \right| ^{2}  \notag \\
&=&\int \frac{d^{2}z_{2}}{\pi }\sec h^{2}\lambda \exp \left( -\left|
z_{2}\right| ^{2}\right) :\exp \left( z_{2}a_{1}tanh\lambda +z_{2}^{\ast
}a_{1}^{\dagger }tanh\lambda -a_{1}^{\dagger }a_{1}\right) :.  \notag \\
&&  \label{Eq7}
\end{eqnarray}%
Use of the formula%
\begin{equation}
\int \frac{d^{2}z}{\pi }\exp \left( -\left| z\right| ^{2}+\sigma z\right)
f\left( z^{\ast }\right) =f\left( \sigma \right) ,  \label{Eq8}
\end{equation}%
we obtain%
\begin{equation}
\rho _{1}^{tms}=\sec h^{2}\lambda :\exp \left( a_{1}^{\dagger
}a_{1}tanh^{2}\lambda \right) e^{-a_{1}^{\dagger }a_{1}}:.  \label{Eq9}
\end{equation}%
Because 
\begin{equation}
\exp \left( a_{1}^{\dagger }a_{1}tanh^{2}\lambda \right) =\sum_{n=0}^{\infty
}\left( a_{1}^{\dagger }a_{1}tanh^{2}\lambda \right) ^{n}/n!,  \label{Eq10}
\end{equation}%
replacing Eq.(\ref{Eq10}) into Eq.(\ref{Eq9}) we have \cite{Fanbook}%
\begin{equation}
\rho _{1}^{tms}=\sec h^{2}\lambda \left\{ \left| 0\right\rangle \left\langle
0\right| +tanh^{2}\lambda \left| 1\right\rangle \left\langle 1\right|
+...+tanh^{2n}\lambda \left| n\right\rangle \left\langle n\right|
+...\right\} .  \label{Eq11}
\end{equation}%
So the entanglement denoted by von Neumann entropy of $\rho _{1}^{tms}$ is%
\begin{eqnarray}
E^{tms} &=&-\frac{1}{cosh^{2}\lambda }\log \left( \frac{1}{\cos h^{2}\lambda 
}\right) -\frac{tanh^{2}\lambda }{cosh^{2}\lambda }\log \left( \frac{%
tanh^{2}\lambda }{cosh^{2}\lambda }\right) -  \notag \\
&&...-\frac{tanh^{2\left( n-1\right) }\lambda }{cosh^{2}\lambda }\log \left( 
\frac{tanh^{2\left( n-1\right) }\lambda }{cosh^{2}\lambda }\right) -... 
\notag \\
&=&\frac{1}{cosh^{2}\lambda }\left( \log \left( cosh^{2}\lambda \right)
-tanh^{2}\lambda \log \left( \sinh ^{2}\lambda \right) \right)   \notag \\
&&\times \left( 1+2tanh^{2}\lambda +...+\left( n+1\right) tanh^{2n}\lambda
+...\right) .  \label{Eq12}
\end{eqnarray}%
From the relationship%
\begin{equation}
1+2tanh^{2}\lambda +...+\left( n+1\right) \tan h^{2n}\lambda
+...=cosh^{4}\lambda ,  \label{Eq13}
\end{equation}%
we have%
\begin{equation}
E^{tms}=cosh^{2}\lambda \log \left( cosh^{2}\lambda \right) -\sinh
^{2}\lambda \log \left( \sinh ^{2}\lambda \right) ,  \label{Eq14}
\end{equation}%
From Eq.(\ref{Eq14}) we can plot $E^{tms}$ with the parameter $\lambda $ as
figure 1%
\begin{eqnarray*}
&& \\
&&figure1 \\
&&
\end{eqnarray*}%
From Fig. 1 we see that the entanglement $E^{tms}$ increase with the
increasing of squeezed parameter $\lambda ,$ which is coincided with the
result obtained by other methods \cite{squeezed}, \cite{squeezed1} \cite%
{squeezed2}.

Secondly, we study the generating of entanglement of free single particle in
thermal environment. In 1975, Takahashi and Umezawa \cite{TFD} pointed out
that the statistical average of physical quantity $Q$ at the temperature T
can be expressed as%
\begin{equation}
\left\langle Q\right\rangle =\left\langle 0\left( \beta \right) \left|
Q\right| 0\left( \beta \right) \right\rangle ,  \label{Eq15}
\end{equation}%
where the thermal quantum vacuum state $\left| 0\left( \beta \right)
\right\rangle $ is defined by%
\begin{equation}
\left| 0\left( \beta \right) \right\rangle =Z\left( \beta \right) ^{-\frac{1%
}{2}}\sum_{n}e^{-\beta E_{n}/2}\left| n,\tilde{n}\right\rangle .
\label{Eq16}
\end{equation}%
Here, we have%
\begin{equation}
Z\left( \beta \right) =Tre^{-\beta \mathcal{H}},\quad \beta =1/k_{B}T,
\label{Eq17}
\end{equation}%
for the relevant Hamiltonian $\mathcal{H}$ and $\left| n\right\rangle $
denotes the $n-th$ eigenstate of the Hamiltonian with the eigenvalue $E_{n}$%
, and the $\left| \tilde{n}\right\rangle $ is the corresponding eigenstate
in the fictitious dynamical system identical to the original system. It is
recently pointed out that the fictitious (tilde) system $\mathcal{\tilde{H}}$
in fact denotes the system of environment \cite{MPLATFD} (in which it is
regarded as one particle). The thermal vacuum state $\left| 0\left( \beta
\right) \right\rangle $ is usually a entangled state and the entanglement is
that between the system of interest and its environment. In the following we
calculate the entanglement of thermal state of a free boson. The Hamiltonian
of free boson is $\mathcal{H}=\omega a^{\dagger }a$ and its thermal vacuum
state is 
\begin{equation}
\left| 0\left( \beta \right) \right\rangle =\left( 1-e^{-\beta \omega
}\right) ^{\frac{1}{2}}\exp (e^{-\frac{\beta \omega }{2}}a^{\dagger }\tilde{a%
}^{\dagger })\left| 0\tilde{0}\right\rangle ,  \label{Eq18}
\end{equation}%
then 
\begin{equation}
\rho _{12}^{tv}=\left( 1-e^{-\beta \omega }\right) \exp (e^{-\frac{\beta
\omega }{2}}a^{\dagger }\tilde{a}^{\dagger })\left| 0\tilde{0}\right\rangle
\left\langle 0\tilde{0}\right| \exp (e^{-\frac{\beta \omega }{2}}a\tilde{a}),
\label{Eq19}
\end{equation}%
where and in the following the superscript ``tv'' denotes thermal vacuum
state. The reduce density matrix of the system (indicated by subscript $1)$
is%
\begin{eqnarray}
\rho _{1}^{tv} &=&Tr_{2}\left\{ \rho _{12}\right\} =Tr_{2}\left\{ \left(
1-e^{-\beta \omega }\right) \exp (e^{-\frac{\beta \omega }{2}}a^{\dagger }%
\tilde{a}^{\dagger })\left| 0\tilde{0}\right\rangle \left\langle 0\tilde{0}%
\right| \exp (e^{-\frac{\beta \omega }{2}}a\tilde{a})\right\} .  \notag \\
&&  \label{Eq20}
\end{eqnarray}%
By using the completeness of coherent state $\left| \tilde{z}\right\rangle $
of the fictitious system%
\begin{equation}
\int \frac{d^{2}\tilde{z}}{\pi }\left| \tilde{z}\right\rangle \left\langle 
\tilde{z}\right| =1,  \label{Eq21}
\end{equation}%
we have%
\begin{eqnarray}
\rho _{1}^{tv} &=&Tr_{2}\{\int \frac{d^{2}\tilde{z}}{\pi }\left| \tilde{z}%
\right\rangle \left\langle \tilde{z}\right| \left( 1-e^{-\beta \omega
}\right) \exp (e^{-\frac{\beta \omega }{2}}a^{\dagger }\tilde{a}^{\dagger
})\left| 0\tilde{0}\right\rangle \left\langle 0\tilde{0}\right| \exp (e^{-%
\frac{\beta \omega }{2}}a\tilde{a})\}  \notag \\
&=&\int \frac{d^{2}\tilde{z}}{\pi }\left\langle \tilde{z}\right| \left(
1-e^{-\beta \omega }\right) \exp (e^{-\frac{\beta \omega }{2}}a^{\dagger }%
\tilde{a}^{\dagger })\left| 0\tilde{0}\right\rangle \left\langle 0\tilde{0}%
\right| \exp (e^{-\frac{\beta \omega }{2}}a\tilde{a})\left| \tilde{z}%
\right\rangle   \notag \\
&=&\left( 1-e^{-\beta \omega }\right) \exp \left( e^{-\beta \omega
}a^{\dagger }a\right) \left| 0\right\rangle \left\langle 0\right|   \notag \\
&=&\left( 1-e^{-\beta \omega }\right) \sum_{n=0}^{\infty }e^{-n\beta \omega
}\left| n\right\rangle \left\langle n\right| .  \label{Eq22}
\end{eqnarray}%
So we can calculate its entanglement as%
\begin{equation}
E^{tv}=-\log \left( 1-e^{-\beta \omega }\right) -\frac{e^{-\beta \omega }}{%
1-e^{-\beta \omega }}\log e^{-\beta \omega }.  \label{Eq23}
\end{equation}%
The entanglement $E^{tv}$ with temperature T plotted in figure 2.%
\begin{eqnarray*}
&& \\
&&figure2 \\
&&
\end{eqnarray*}%
It is shown that when $T\rightarrow 0,$ $E^{tv}\rightarrow 0$ which shows
that the free single particle do not entangle with its environment at very
low temperature and when the temperature increase the entanglement emerges.
In the room temperature the entanglement increase with the increasing of
temperature. This result is physically reasonable.

Thirdly, we calculate the entanglement of two coupling harmonic oscillators.
This system can model some molecules through modeling atoms with harmonic
oscillators. The simplest form of two coupling harmonic oscillator system
has the Hamiltonian as%
\begin{equation}
\mathcal{H}=\frac{1}{2m}\left( p_{1}^{2}+p_{2}^{2}\right) +\frac{1}{2}%
m\omega ^{2}\left( x_{1}^{2}+x_{2}^{2}\right) -\delta x_{1}x_{2}.
\label{Eq24}
\end{equation}%
Here, $m$ denotes the mass and the $\omega $ denotes the frequency of
particles (suppose they are same for the two particles), $\delta $ denotes
the coupling constant and 
\begin{equation}
x_{i}=\frac{a_{i}+a_{i}^{\dagger }}{\sqrt{2}},\quad p_{i}=\frac{%
a_{i}-a_{i}^{\dagger }}{j\sqrt{2}},(i=1,2,j^{2}=-1).  \label{Eq25}
\end{equation}%
When this system acting on the vacuum state, the eigenstate of energy is $%
U\left| 00\right\rangle $ where \cite{twoatom} 
\begin{equation}
U=\left( \frac{\omega ^{2}}{\omega _{1}\omega _{2}}\right) ^{\frac{1}{4}%
}\int \int\limits_{-\infty }^{\infty }dx_{1}dx_{2}\left| u\left( 
\begin{array}{c}
x_{1} \\ 
x_{2}%
\end{array}%
\right) \right\rangle \left\langle \left( 
\begin{array}{c}
x_{1} \\ 
x_{2}%
\end{array}%
\right) \right| .  \label{Eq26}
\end{equation}%
Here,%
\begin{equation*}
\omega _{1}^{2}=\omega ^{2}-\frac{\delta }{m},\quad \omega _{1}^{2}=\omega
^{2}-\frac{\delta }{m},
\end{equation*}%
\begin{equation}
u=\left( 
\begin{array}{cc}
\sqrt{\frac{\omega }{2\omega _{1}}} & \sqrt{\frac{\omega }{2\omega _{2}}} \\ 
\sqrt{\frac{\omega }{2\omega _{1}}} & -\sqrt{\frac{\omega }{2\omega _{2}}}%
\end{array}%
\right) ,\quad \det u=\frac{\omega }{\sqrt{\omega _{1}\omega _{2}}}.
\label{Eq27}
\end{equation}%
By using IWOP one can obtain \cite{twoatom} 
\begin{eqnarray}
U &=&\left( \sec hr_{1}\sec hr_{2}\right) ^{\frac{1}{2}}\exp \left\{ \frac{1%
}{4}\left( a_{1}^{\dagger 2}+a_{2}^{\dagger 2}\right) ^{2}tanhr_{1}+\frac{1}{%
4}\left( a_{1}^{\dagger 2}-a_{2}^{\dagger 2}\right) ^{2}tanhr_{2}\right\}  
\notag \\
&:&\exp \left\{ \left( \frac{1}{\sqrt{2}}\sec hr_{1}-1\right) a_{1}^{\dagger
}a_{1}+\frac{1}{\sqrt{2}}a_{2}^{\dagger }a_{1}\sec hr_{1}-\frac{a_{1}^{2}}{2}%
tanhr_{1}\right\} :  \notag \\
&:&\exp \left\{ \left( -\left( \frac{1}{\sqrt{2}}\sec hr_{2}+1\right)
a_{2}^{\dagger }a_{2}-\frac{1}{\sqrt{2}}a_{1}^{\dagger }a_{2}\sec hr_{2}-%
\frac{a_{2}^{2}}{2}tanhr_{2}\right) \right\} :,  \notag \\
&&  \label{Eq28}
\end{eqnarray}%
where%
\begin{equation}
tanhr_{i}=\frac{\omega -\omega _{i}}{\omega +\omega _{i}},\quad \sec hr_{i}=%
\frac{2\sqrt{\omega \omega _{i}}}{\omega +\omega _{i}},i=1,2.  \label{Eq29}
\end{equation}%
Thus, the density matrix under the operator $U$ acting on vacuum state is%
\begin{eqnarray}
\rho _{12}^{cho} &=&\sec hr_{1}\sec hr_{2}\exp \left\{ \frac{1}{2}\left(
a_{1}^{\dagger 2}tanhr_{1}+a_{2}^{\dagger 2}tanhr_{2}\right) \right\} \left|
00\right\rangle \left\langle 00\right|   \notag \\
&&\exp \left\{ \frac{1}{2}\left(
a_{1}^{2}tanhr_{1}+a_{2}^{2}tanhr_{2}\right) \right\} ,  \label{Eq30}
\end{eqnarray}%
where and in the following the superscript ``cho'' denotes coupling harmonic
oscillators. So we have%
\begin{eqnarray}
\rho _{1}^{cho} &=&\sec hr_{1}:\exp \left\{ \frac{1}{2}\left( a_{1}^{\dagger
2}+a_{1}^{2}\right) tanhr_{1}\right\} e^{-a_{1}^{\dagger }a_{1}}:  \notag \\
&=&\sec hr_{1}\sum_{n,m=0}\frac{\left( \frac{1}{2}tanhr_{1}\right) ^{m+n}}{%
m!n!}a_{1}^{\dagger 2n}\left| 0\right\rangle \left\langle 0\right|
a_{1}^{2m}.  \label{Eq31}
\end{eqnarray}%
Suppose the temperature of the two-atom molecule is very low, there are only
two level in the atoms then the density matrix becomes%
\begin{equation}
\rho _{1,1}^{cho}=\sec hr_{1}\left( 
\begin{array}{ccc}
1 & 0 & \frac{1}{\sqrt{2}}tanhr_{1} \\ 
0 & 0 & 0 \\ 
\frac{1}{\sqrt{2}}tanhr_{1} & 0 & \frac{1}{2}tanh^{2}r_{1}%
\end{array}%
\right) .  \label{Eq32}
\end{equation}%
From this density matrix we can calculate the entropy entanglement of this
system, it is%
\begin{equation}
E_{1}^{cho}=-(sechr_{1}+\frac{1}{2}sechr_{1}tanh^{2}r_{1})\log (sechr_{1}+%
\frac{1}{2}sechr_{1}tanh^{2}r_{1}).  \label{Eq33}
\end{equation}%
From Eq.(\ref{Eq33}) we can plot $E_{1}^{cho}$ with squeezed parameter $r_{1}
$ as figure 3. 
\begin{eqnarray*}
&& \\
&&figure3 \\
&&
\end{eqnarray*}%
It is shown that the entanglement take its maximum at $r_{1}\approx 2.$ A
furthermore calculation shows that when the temperature raise and the level
of each harmonic oscillators more than two, for example $n$, the density
matrix $\rho _{1}^{cho}$ becomes a matrix of $\left( 2n-1\right) \times
\left( 2n-1\right) .$ When $n$ big enough, this problem will become very
difficult. So, in order to calculate the entanglement of this system a
better method is expected.

In this paper, by using the results of normally ordered forms of operators
and the IWOP of operators we calculated the generating of entanglement from
biparticle Bose systems acting on vacuum state. These systems include
two-mode squeezed system, free single particle in the thermal environment,
and two coupling harmonic oscillators system. The generating of entanglement
by two mode squeezed operator acting on vacuum state is equal to the
entanglement of two-mode squeezed vacuum state. Our result is coincided with
that has been obtained by many authors. The entanglement produced from free
single particle in thermal environment is one between the particle and the
environment, which is obtained rigorously by using our method. The
entanglement of coupled two harmonic oscillators is the entanglement between
one particle and the other. By using our result one can obtain the
approximate degree of entanglement of this system at arbitrarily accurate
degree, theoretically. Of course, it is difficult when these particles in a
higher temperature. It is shown that IWOP is a useful tool for the
investigating of generating of entanglement on biparticle Bose systems.

\begin{acknowledgement}
This project partly supported by Scientific Research Fund of Hunan
Provincial Education Department under Grand No. 01C036. I thank professor
Hong-Yi Fan for his guidance.
\end{acknowledgement}

\section{The captious of the figures}

Figure 1: $E^{tms}$ with squeezed parameter $\lambda .$

Figure 2: $E^{tv}$ with teperature $T$ where $\omega =10^{-20},$ $%
k_{B}=1.3806568\times 10^{-23}\unit{J}\unit{K}^{-1}.$

Figure 3: $E_{1}^{cho}$ with the squeezed parameter $r_{1}.$

\end{document}